# SINOMA - A new approach for estimating linear relationships between noisy serial data streams

Barnim Thees[1], Allan Buras[2*], Gottfried Jetschke[3], Eduardo Zorita[4], Martin Wilmking[2], Volkmar Liebscher[5], and Lars Kutzbach[1]

[1] Institute of Soil Science, Center for Earth System Research and Sustainability, Universität Hamburg, Allende-Platz 2, 20146 Hamburg.

[2] Institute of Botany and Landscape Ecology, Ernst-Moritz-Arndt Universität, Greifswald, Soldmannstraße 15, 17487 Greifswald.

[3] Institute of Ecology, University of Jena, Dornburger Straße 159, 07743 Jena.

[4] Institute of Coastal Research, Helmholtz-Zentrum Geesthacht, Max-Planck-Straße 1, 21502 Geesthacht.

[5] Institute of Mathematics and Informatics, Ernst-Moritz-Arndt Universität, Greifswald, Walter-Rathenau-Straße 47, 17487 Greifswald.

* corresponding author: Allan Buras; e-mail: allan@buras.eu Phone: +493834864185, Fax: +493834864114

**Abstract**

Reconstructions of past climates are based on the calibration of available proxy data. This calibration is usually achieved by means of linear regression models. In the recent paleo-climate literature there is an ongoing discussion on the validity of highly resolved (annual time scales) climate reconstructions. The reason for this is that the proxy data are noisy, i.e. in addition to the variability that is related to the climate variable of interest, they contain other sources of variability. Inadequate treatment of such noise leads to a biased estimation of regression slopes, resulting in a wrong representation of the real amplitude of past climate variations. Methods to overcome this problem have had a limited success so far. Here, we present a new approach – SINOMA – for noisy serial data streams (e.g. time series, spatial transects) that are characterized by different spectral characteristics of signal and noise. SINOMA makes use of specific properties of the data streams' temporal/spatial structure and by this is able to deliver a precise estimate of the true regression slope *and, simultaneously*, of the ratio of noise variances present in the predictor and the predictand. The paper introduces the underlying mathematics as well as a general description of the presented algorithm. The validity of SINOMA is illustrated with two test data sets. Finally, we address methodological limitations and further potential applications.

# 1     General introduction

Estimating a linear relationship between two variables that are both disturbed by noise is a common problem in many scientific fields, e.g. in climate sciences (e.g. Allen and Stott, 2003), ecology (e.g., McArdle, 1988; Carpenter *et al.*, 1994), econometrics (e.g., Hausman, 2001), or astronomy (e.g., Akritas and Bershady, 1996). Particularly, the issue has been vigorously discussed over the last decade in the context of paleo-climate reconstruction methodology (e.g., Esper et al., 2002; Zorita et al., 2003; von Storch et al., 2004; Moberg et al., 2005; Hegerl et al., 2007; Riedwyl et al., 2009; Christiansen et al., 2009; Tingley and Huybers, 2010a,b; Christiansen 2011; Christiansen and Ljungqvist, 2011; Kutzbach et al., 2011; Moberg and Brattström, 2011; Christiansen 2012; Christiansen and Ljungqvist, 2012a,b; Moberg, 2012; Tingley and Li, 2012; Tingley et al., 2012). For reasons of simplicity, we will focus the discussion of the presented new approach on its implications for the reconstruction of paleo-climatic time series. Nevertheless, the new methodology will also be applicable for many other scientific fields dealing with relationships between noisy serial data streams (e.g. also spatial transects).

For the non-climatologist reader, a short introduction into the paleo-climate reconstruction problem shall be given here: The reconstruction of past climates is based on the assumption that natural environmental archives – like tree-rings, coral skeletons or ocean and lake sediments – reach back beyond instrumental records and thus contain information that can be used as proxies ("approximations of") for certain target climatic quantities (e.g. summer air temperature) that influenced the proxy during its formation. Although the relation between the proxy and the target climatic quantity might be complex, the simplifying assumption of a stationary linear relationship between the proxy and the climatic target quantity is commonly adopted (Sachs et al., 1977; Fritts et al., 1990; National Research Council, 2006; Kutzbach et al., 2011; Moberg and Brattström, 2011). Accordingly, the effects of secondary environmental controls and disturbances have to be minimized by careful site and tree selection, so that their overall effect can be characterized as white noise disturbing the effect of the target climatic variable. The linear relationship is then used to predict (i.e. reconstruct) the target climatic quantity over the past using the proxy time series as predictor variable. However, despite the multitude of published paleo-climate reconstruction studies, the central question is still controversially debated: How to best estimate/calibrate the parameters of the linear model that describes the relationship between the proxy and the target climatic quantity (e.g., National Research Council, 2006; Christiansen, 2011; 2012; Christiansen and Ljungqvist, 2011, 2012a,b; Moberg, 2012; Tingley et al., 2012).

Direct ordinary least squares (OLS) regression with the proxy as the predictor variable and the target climatic quantity as the predictand variable has traditionally been the most often applied approach (National Research Council, 2006). But it is now widely acknowledged that direct OLS regression – due to the basic assumption of a noise-free proxy variable – will lead to biased model parameter estimates and underestimated amplitudes of the reconstructed climate time series at all frequencies (e.g., Hegerl *et al.*, 2007; Riedwyl *et al.*, 2009; Christiansen and Ljungqvist, 2011; Moberg and Brattström, 2011). Some authors argue for inverse OLS (INV) regression, thereby neglecting the noise in the target climatic quantity time series (Coelho et al, 2004; Christiansen, 2011, 2012; Christiansen and Ljungqvist, 2011, 2012a,b). In this context, we argue that both the proxy time series *and* the target climatic quantity time series have to be considered as contaminated with noise (Jones et al., 1997; Brohan et al., 2006) agreeing in this aspect with e.g. Ammann et al. (2010), Moberg and Brattström (2011), Moberg (2012), and Tingley and Li (2012). If the target climatic quantity is noisy, the INV model would lead to overestimated amplitudes at all frequencies (Kutzbach et al., 2011; Moberg and Brattström, 2011; Tingley and Li, 2012).

A solution for this more general case where both predictor and predictand measurements contain noise is known as the 'Errors-In-Variables' model (EVM) (e.g., Fuller, 1987; Cheng and van Ness, 1999). Although its application should be very common, this model is in practice rarely applied within the paleo-climatology community (but see e.g. Hegerl et al., 2007; Amman et al., 2010; Moberg and Brattström, 2011). The main constraint of the EVM approach is that information on either the noise variance of the predictand variable, the noise variance of the predictor variable, or on the ratio of these noise variances is needed, which seldom is available (Kutzbach et al., 2011; Christiansen 2011, 2012; Moberg and Brattström, 2011).

The early statistical literature has already shown that the parameters of the EVM model are not identifiable without *a priori* information if the predictor and predictand variables and the noise processes are jointly normal independent random variables with zero expected value and unknown variances (Frisch, 1934; Kendall and Stuart, 1973; Söderström, 2007). However, if the predictor and predictand variables have non-normal distributions (Reiersøl, 1950; Deistler and Anderson, 1989) and/or are characterized by serial correlation (Mehra, 1976; Söderström, 1980, 2007; Solo, 1986, Castaldi and Soverini, 1996; Agüero and Goodwin, 2008), as is the case in many of the applications mentioned above, the model theoretically is identifiable.

With this paper, we introduce the **S**equential **I**terative **NO**ise **M**atching **A**lgorithm (SINOMA) which recently has been designed to derive consistent estimates of the parameters of a linear relationship between two noisy time series and of their respective error variances. SINOMA makes use of the fact that the undisturbed predictor and predictand variables in many applications of interest are time series that have a fixed order in time, serial correlation and low-frequency fluctuations that dominate the power spectrum. On the other hand, we assume that the noise processes that disturb the variables are approximately white. Therefore, the spectral properties of the undisturbed variables and the noise processes, respectively, are assumed to be different. Two time-varying variables which have a linear relationship are expected to have parallel low-frequency fluctuations of proportional amplitudes. This feature allows for separation of the co-varying undisturbed variables from the disturbing, statistically independent white noise processes of these time series when analyzed together. While others used spectral techniques to analyze the identifiability of dynamic EVM models (e.g., Söderström, 1980; Anderson and Deistler, 1984; Stoica and Nehorai, 1987), SINOMA is working in the time domain.

SINOMA splits the two noisy time series into short sequences which are defined by the order of local extreme values and for which local means and variances can be estimated. The variability of the local means is dominated by the lower-frequency components of the undisturbed, noise-free variables, whereas the local variances are composed of the high-frequency fluctuations of the disturbing noise and the undisturbed variables, respectively. SINOMA relies on the assumption that the highest-frequency fluctuations of the time series – and therefore the definition of the sequences with their local variances – are dominated by white noise. SINOMA is based on a group of structural similarity indices called *Explanatory Powers* (EP) that have been designed to simultaneously evaluate to which extent two time series on average have equal local means and equal local variances. With these EP indices, it becomes possible to evaluate the ratio of the error variances of the two noisy serial data streams. Specifically, the EP indices allow to estimate whether the error variances conform with reduced major axis (RMA) regression conditions (i.e., ratio of the error variances ($\lambda$) equals the square of the slope, thus $\lambda = c^2$) or deviates from RMA conditions in the direction of OLS conditions ($\lambda \rightarrow \infty$) or in the direction of INV conditions ($\lambda \rightarrow 0$). SINOMA iteratively adds numerically realized white noise of varying variance magnitude to both of the two serial data streams until valid error conditions for RMA regression are reached. When RMA conditions are met, the consistent estimates of the model parameters (offset and slope) of the linear relationship between the non-noisy variables can be calculated e.g. by applying the RMA regression. Furthermore, the error variances of the noise processes originally disturbing the predictand and predictor variables, respectively, are identifiable.

In section 2 we will give a short review on general model approaches and introduce three datasets that shall accompany the reader along the theoretical explanations of SINOMA. Section 3 introduces the *Explanatory Powers* EP indices. Section 4 describes the features of the EP indices within the range of EVM and how these are used to identify RMA conditions, which is the central procedure within SINOMA. In section 5 SINOMA is validated on a pseudo-proxy dataset. Finally, in section 6 we discuss the applicability, constraints and possible future applications of SINOMA. An overview on the nomenclature used throughout the manuscript is given in the Appendix (Table A1).

**2. OLS, INV, RMA and general EVM conditions**

**2.1 Theoretical background and terminology**

For simplicity, our paper will focus on a linear one-dimensional model between the predictor variable *x* and the predictand variable *y* where the unknown parameters $c$ and $c_0$ are to be determined:

$$y = c \cdot x + c_0 \tag{1a}$$

The predictor variable *x* and the predictand variable *y* are both time series that have a fixed order in time and serial correlation, i.e. they have spectral characteristics deviating from white noise. Both variables are not directly observable. Their observations are prone to errors leading to the series *x′* and *y′* which are disturbed by pair-wise independent noise terms $\varepsilon$ and $\delta$:

$$x' = x + \varepsilon \quad , y' = y + \delta \tag{1b}$$

Here, we suppose that both time series $\varepsilon$ and $\delta$ are independent temporal sequences of identically distributed random variables with zero mean and series-specific constant variances $S_\varepsilon$ and $S_\delta$, respectively. Consequently, both series represent perfect time-discrete white noise processes. Furthermore, we suppose that the errors $\varepsilon$ and $\delta$ should be uncorrelated with the true values x and y:

$$E(\varepsilon_i) = E(\delta_i) = 0, \quad \operatorname{var} \varepsilon_i = S_\varepsilon^2, \quad \operatorname{var} \delta_i = S_\delta^2, \quad \text{for all } i = 1,\dots,N \tag{2a}$$

$$\operatorname{cov}(\varepsilon_i, \varepsilon_j) = \operatorname{cov}(\delta_i, \delta_j) = 0, \quad i \neq j, . \tag{2b}$$

$$\operatorname{cov}(\varepsilon_i, \delta_j) = \operatorname{cov}(x_i, \varepsilon_j) = \operatorname{cov}(x_i, \delta_j) = \operatorname{cov}(y_i, \varepsilon_j) = \operatorname{cov}(y_i, \delta_j) = 0 \quad \text{all } i,j \tag{2c}$$

For finite samples these assumptions will however not be ideally fulfilled and in reality correlations between noise and signal of various strengths are likely to occur. These effects will of course influence the accuracy of SINOMA as they would do for all regression models (but see section 6.2).

Since we assume a linear relationship between the real values $x$ and $y$, the predicted values $\hat{y}'$ are defined through estimates $\hat{c}$ and $\hat{c}_0$ as a linear function of observations $x'$:

$$\hat{y}' = \hat{c} \cdot x' + \hat{c}_0 \tag{3}$$

According to our assumptions about the error noise we have

$$\bar{x}' = \bar{x} \quad , \bar{y}' = \bar{y} = c \cdot \bar{x} + c_0 \quad , \bar{\hat{y}}' = \hat{c} \cdot \bar{x}' + \hat{c}_0 \text{ and} \tag{4a}$$

$$S_{x'}^2 = S_x^2 + S_\varepsilon^2 \quad , S_{y'}^2 = S_y^2 + S_\delta^2, \quad S_{x'y'} = c \cdot S_x^2 = c \cdot (S_{x'}^2 - S_\varepsilon^2) \tag{4b}$$

where empirical means and variances are calculated for the corresponding time series. If the noise variance in either the *x'*- or *y'*- variable is known, the resulting EVM (= Errors-in-Variables Model) slope can be achieved by:

$$\hat{c}_{EVM} = \frac{S_{x'y'}}{S_{x'}^2 - S_\varepsilon^{S2}} = \frac{S_{y'}^2 - S_\delta^2}{S_{x'y'}} \tag{5}$$

However, if the noise variances are not known, equation (5) cannot be used. Only under additional assumptions explicit analytical estimates can be derived for the unknown parameters $c$ and $c_0$ (slope and intercept, see e.g. Fuller, 1987, Hartung, 1999, Kutzbach et al., 2011). The most important case is the situation where the noise ratio $\lambda$ is known with

$$\lambda := \frac{S_\delta^2}{S_\varepsilon^2}, \tag{6}$$

Slope and intercept are then given by the general EVM formula

$$\hat{c}_{EVM} = \frac{S_{y'}^2 - \lambda \cdot S_{x'}^2}{2 \cdot S_{x'y'}} + \frac{\sqrt{(S_{y'}^2 - \lambda \cdot S_{x'}^2)^2 + 4 \cdot \lambda \cdot (S_{x'y'})^2}}{2 \cdot S_{x'y'}} =: \hat{c}_{EVM}(\lambda) \tag{7a}$$

$$\hat{c}_{0_{EVM}} = \bar{y}' - \hat{c}_{EVM} \cdot \bar{x}', \tag{7b}$$

where the slope is the larger solution of the quadratic equation

$$S_{x'y'} \cdot \hat{c}_{EVM}^2 + (\lambda \cdot S_{x'}^2 - S_{y'}^2) \cdot \hat{c}_{EVM} - \lambda \cdot S_{x'y'} = 0. \tag{8}$$

In this manuscript, $\hat{c}_{EVM}$ is always related to equation (7a) and considered as a function of $\lambda$. Through the rest of the paper it will be convenient to assume (without loss of generality) that the slope is always positive, hence $S_{x'y'} > 0$.

Three special cases exist for which the expression of equation (7a) simplifies remarkably to common approaches mentioned in the literature (see Table 1):

(a) For noiseless $x' = x$ data $\lambda$ tends to infinity and (7a) simplifies to the Ordinary Least Squares model (OLS, the most frequently considered assumption in the literature):

$$\hat{c}_{OLS} = \frac{S_{xy'}}{S_x^2} \tag{9a}$$

(b) If $\lambda$ equals the square of the slope $c$, then $S_y/S_x = S_{y'}/S_{x'}$ and (7a) simplifies to the Reduced Major Axis model (RMA, aka variance matching model):

$$\hat{c}_{RMA} = \sqrt{\lambda_{RMA}} = \frac{S_{y'}}{S_{x'}} \tag{9b}$$

(c) For noiseless $y' = y$ data $\lambda$ becomes zero, and (7a) simplifies to the inverse OLS model (INV):

$$\hat{c}_{INV} = \frac{S_y^2}{S_{x'y}} \tag{9c}$$

Referring to these definitions, data will be called 'between OLS and RMA' for $S_{y'}^2/S_{x'}^2 < S_\delta^2/S_\varepsilon^2$ and 'between RMA and INV' for $S_{y'}^2/S_{x'}^2 > S_\delta^2/S_\varepsilon^2$. From the well-known relation $S_{x'y'} \leq S_{x'} \cdot S_{y'}$ it can be easily derived that the inequality

$$\hat{c}_{OLS} \leq \hat{c}_{RMA} \leq \hat{c}_{INV} \tag{10}$$

holds (please see Table 1 and Kutzbach *et al.* (2011) for details). This means that if both x and y are noisy the OLS model will underestimate the true slope, the INV model will overestimate the true slope for noise in *x and y* data, and the RMA model will over- or underestimate the true slope if $\lambda$ is smaller or larger than $S_{y'}^2/S_{x'}^2$, respectively.

Table 1: Three special solutions of equation (7a) based on either OLS, INV or RMA assumptions regarding the type of error noise in the data, and the respective variance ratio of model and observation.

| model | OLS | RMA | INV |
|---|---|---|---|
| valid for noise ratio $\lambda =$ | $\infty$ | $S_{y'}^2/S_{x'}^2$ | 0 |
| estimated slope $\hat{c}_{EVM}(\lambda)$ | $\hat{c}_{OLS} = S_{xy'}/S_x^2$ | $\hat{c}_{RMA} = S_{y'}/S_{x'}$ | $\hat{c}_{INV} = S_y^2/S_{x'y}$ |
| variance ratio $S_{\hat{y}'_{EVM}}^2/S_{y'}^2$ | $R^2$ | 1 | $1/R^2$ |

In most cases, there is no knowledge on the single noise intensities or about the noise ratio $\lambda$ which is needed to calculate equation (7a). So far, the common procedure to treat such datasets was to either assume OLS, RMA, or INV data and accordingly calculate (9a-c). Modeled time series data $\hat{y}'$ obtained by these approaches will be denoted 'OLS model', 'RMA model' and 'INV model' data, respectively. As shown, the parameters of these models are only consistent for particular noise ratios $\lambda$, while the estimate of $\hat{c}_{EVM}$ from equation (7a), for known $\lambda$ is always a consistent estimator of the true slope (Kutzbach et al., 2011).

Using the Pearson correlation coefficient $R$ between $x'$ and $y'$, the following relations between (9a-c) will be useful in later calculations:

$$\frac{1}{R} \cdot \hat{c}_{OLS} = \hat{c}_{RMA} = R \cdot \hat{c}_{INV} \tag{11}$$

From (11) we immediately see that $\hat{c}_{INV} = \hat{c}_{OLS}/R^2$, hence the maximal underestimation of the real slope occurs for the OLS model under INV conditions and equals $R^2$. In contrast, the maximal overestimation (INV model in case of OLS conditions) is $1/R^2$. However, the real error in slope estimates depends on how the noise, contributing to a certain $R$ value, is distributed between both series.

Equation (8) can uniquely be solved for $\lambda$ as a function of the slope estimate $\hat{c}$ because $\hat{c}(\lambda)$ is strictly decreasing in $\lambda$. Interestingly, this solution may be expressed in a rather simple form by using the particular model slope estimates from (9a-c) and the Pearson correlation coefficient $R$ as abbreviations:

$$\lambda = \frac{\hat{c}_{INV} - \hat{c}_{EVM}}{\left(\frac{1}{\hat{c}_{OLS}} - \frac{1}{\hat{c}_{EVM}}\right)} = \frac{\frac{\hat{c}_{RMA}}{R} - \hat{c}_{EVM}}{\left(\frac{1}{\hat{c}_{OLS}} - \frac{1}{\hat{c}_{EVM}}\right)} = \frac{\frac{\hat{c}_{OLS}}{R^2} - \hat{c}_{EVM}}{\left(\frac{1}{\hat{c}_{OLS}} - \frac{1}{\hat{c}_{EVM}}\right)} \tag{12}$$

This property will be of interest in section 4.

For conceptual reasons we will keep the dash in the symbol $\hat{y}'_i$ of predicted value to emphasize that our predictions (3) are based on the noisy values $x'_i$ although we will not use the estimated values $\hat{y}_i = \hat{c} \cdot x_i + \hat{c}_0$ calculated at the true values $x_i$. It should be noted that since $S^2_{\hat{y}'} = \hat{c}^2 \cdot S^2_{x'}$ (as a consequence of (4a), see (A2) in appendix AI), the variance ratio of estimated $\hat{y}'_i$ (based on proper $\lambda$) and observed $y'_i$ can easily be calculated under general EVM conditions and written as (cf. Table 1 and appendix AI for deductions of equations 9-13):

$$\frac{S^2_{\hat{y}'EVM}}{S^2_{y'}} = \hat{c}^2_{EVM} \cdot \frac{S^2_{x'}}{S^2_{y'}} = \hat{c}^2_{EVM} \cdot \frac{R^2}{\hat{c}^2_{OLS}} \quad . \tag{13}$$

## 2.2 Example datasets

To visualize the described mathematical relationships, we introduce three datasets, which represent almost OLS, RMA and INV conditions. The true $x$ data were generated as one complete period of a time-discrete sine function, while the true $y$ data were defined as the same sine function multiplied with the particular constant $c = 2.1$ (i.e. intercept zero):

$$x_i := \sin\frac{2\pi i}{N} \quad , \quad y_i := c \cdot x_i = c \cdot \sin\frac{2\pi i}{N} \quad , \quad i = 1,\ldots,128 \quad c = 2.1 \tag{14}$$

The time series of observations $x'_i = x_i + \varepsilon_i$, $y'_i = y_i + \delta_i$ used later on were obtained by adding independent, uniformly distributed random error values $\varepsilon$ and $\delta$ of known variances $S_\varepsilon^2$ and $S_\delta^2$ and zero mean (i.e. 'white noise', Shumway and Stoffer, 2011). Random number generation was done according to the 'Mersenne-Twister' algorithm (Matsumoto and Nishimura, 1998). The corresponding $\lambda$ were defined to meet nearly OLS, RMA or INV conditions. Particular numerical realizations of these three cases are visualized in Fig. 1 while the corresponding numerical values are given in Table 2. Since the structure of the noise in these datasets is known, it enables us to test the applicability of our methods introduced later. All calculations based on these datasets were done in 'R' (Version 3.0.1, R Foundation for Statistical Computing, Vienna, Austria).

Table 2: Noise ratios used in the introduced model datasets and empirical sample estimates of $R^2$. For technical reasons of the method introduced later a very small noise was added to $x$ in OLS conditions and to $y$ in INV conditions.

| conditions | OLS | RMA | INV |
|---|---|---|---|
| $S_\delta^2$ | 2.2 | 0.860 | 0.00022 |
| $S_\varepsilon^2$ | 0.00005 | 0.195 | 0.5 |
| $\lambda = S_\delta^2/S_\varepsilon^2$ | 44100 | 4.41 (= 2.1•2.1) | 0.00044 |
| empirical $R^2$ | 0.49 | 0.50 | 0.50 |

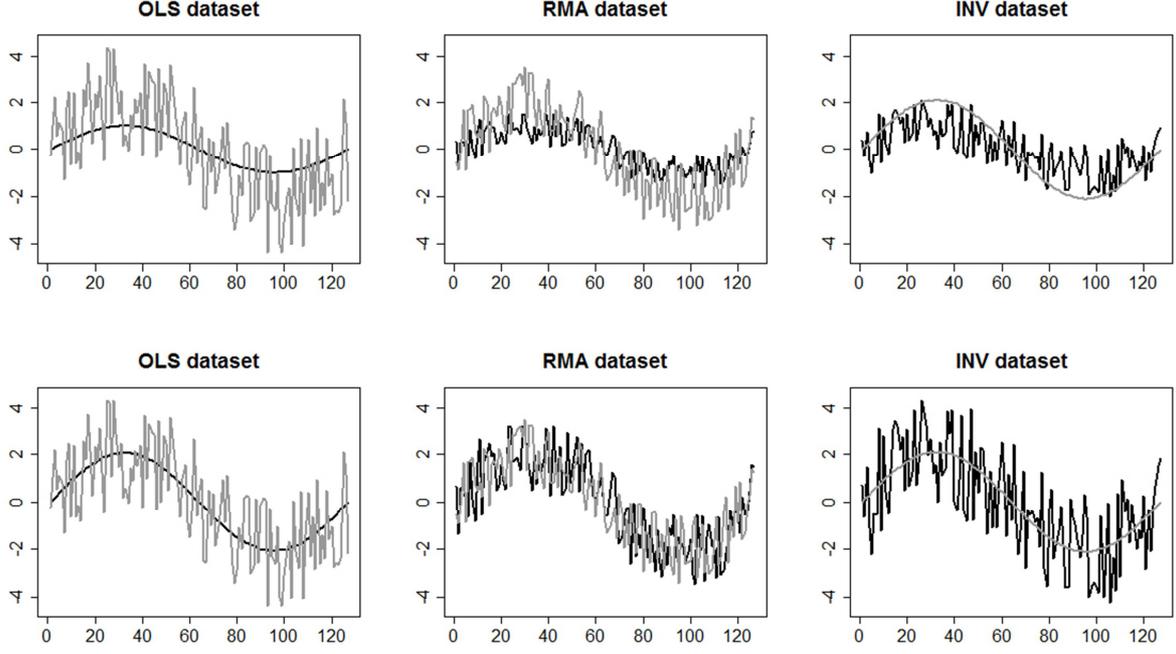

**Fig. 1** upper panels: The three introduced sinusoidal model datasets representing either OLS, RMA, or INV conditions: $x'$ data are plotted in black, $y'$ data in grey. For details on the data properties please see Table 2. Lower panel: correct EVM model estimates $\hat{y}'$ (black), achieved from equation (3) with the model parameter estimates ($c_0$ and $c$) derived from (7), plotted together with measured $y'$ (grey) for the three example datasets

From Table 3 it can be seen that with decreasing noise ratio $\lambda$ (i.e. increasing noise in $x$) the slope estimate $\hat{c}_{OLS}$ achieved by equation (9a) strictly decreases whereas the variance ratio $S^2_{\hat{y}'_{EVM}}/S^2_{y'}$ from (13) strictly increases. Further, the variance of best estimated model data derived from (7a) is smaller, equal to, or larger than the variance of the measured data for OLS, RMA, and INV data, respectively. Thus, in case of OLS conditions, the modeled $\hat{y}'$ data will give the impression of "underestimating" the variability of the measured $y'$ data. However, in this particular case, they truly represent the correct linear transformation of the noiseless measured $x$ data. Contrary for INV data, the INV-modeled $\hat{y}'$ data appear to "overestimate" the variability of the measured y data because they are obtained as a linear transformation of noisy $x'$ data. In this case, the true slope is estimated by the INV equation as well as in the OLS case (~2.1), but the noise in the $x'$ data (which is absent in $x'$ for the OLS case) then affects the model estimate $\hat{y}'$ leading to the observed high variability. Finally, in the case of RMA data the EVM-modeled $\hat{y}'$ data reproduce the noisy measured $y'$ (but not the noiseless $y$) data very well (Fig. 1 lower panels).

Table 3: Slope estimates from equations (7a) and (9a-c) (OLS, RMA, INV and general EVM model) for the three example datasets. The commonly applied model estimates (i.e. OLS, RMA and INV) are close to the true slope (2.1) only if the noise-ratio matches the requirements of the respective models. The general EVM model estimates with known $\lambda$ in equation (7a) are always close to the true slope.

|  | OLS-data | RMA-data | INV-data |
|---|---|---|---|
| $\hat{c}_{OLS}$ formula | **2.07** | (1.50) | (1.08) |
| $\hat{c}_{RMA}$ formula | (2.95) | **2.11** | (1.48) |
| $\hat{c}_{INV}$ formula | (4.20) | (2.97) | **2.08** |
| $\hat{c}_{EVM}(\lambda)$ formula | **2.07** | **2.10** | **2.07** |
| $S^2_{\hat{y}'_{EVM}}/S^2_{y'}$ | 0.49 | 1.00 | 1.97 |
| range of $\hat{y}'_{EVM}, y'$ | (-2.1, 2.1) , (-4.4, 4.3) | (-3.5, 3.2) , (-3.4, 3.5) | (-4.3, 4.3) , (-2.1, 2.1) |

## 3. The new approach: Structural comparisons of sequential data

### 3.1 General description

The new approach to EVM which we introduce by this paper aims at the comparison of structural properties of the two time series (or more general, sequential data): series $y' \equiv (y'_i)_{i=1}^N$ of observed values (i.e. response variable) and series $\hat{y}' \equiv (\hat{y}'_i)_{i=1}^N$ of predicted response values based on observations $x'$ and slope $\hat{c}_{EVM}$ and intersection $\hat{c}_{0EVM}$ estimates achieved by particular solutions of (7a) and (7b). An important prerequisite of the underlying noiseless $x$ and $y$ data is that their spectral characteristics is different from pure white noise, i.e. not all of their periodic components are present with equal amplitude (Shumway and Stoffer, 2011). Basic structural properties are, among others, the particular temporal order of the data pairs ($y'_i, \hat{y}'_i$) and the 'local' means and 'local' variances, calculated for suitably small time intervals (later called 'elementary fluctuations') around an arbitrary target point i in time. From this we define 'areas of equi-amplitude' for the *y'* and *ŷ'* data and calculate the average local overlap between both series as a measure of congruence (see also Thees et al., 2009). By iterative changes of the properties of the two observed time series *x'* and *y'* (achieved by addition of extra noise) and, consequently, of the modeled time series *ŷ'*, we are able to arrive at a special situation with a maximum of structural similarity of both series, resembling RMA conditions. This allows us to specify both an unbiased estimate of the true slope ĉ as well as of the variances of the original noise components in both series (and thus *λ*).

## 3.2 Sequential structures of 'elementary fluctuations'

In our approach, the local oscillations of $\hat{y}'_i$ and $y'_i$ series are split into sequences of so-called 'elementary fluctuations' (indexed by s). Each fluctuation's range is defined from one local maximum to the next (or minimum to minimum, if necessary) and, therefore, consists of at least $N_S \geq 3$ sequential values of the time series (see Fig. 2, left). For both series the arrangement of all local maxima leads to a specific partitioning of the discrete time axis i = 1, ... , N into a finite number of *M'* (or $\hat{M}$, respectively) intervals. By definition, the i-values of local maxima themselves belong to two neighboring elementary fluctuations.

For each of the fluctuations (s = 1, ..., *M'*) we define a symmetric interval $I'_s$ around the local fluctuation's mean $\bar{y}'_s$ with a vertical length – called bandwidth – $A_s(y') \equiv A'_s$ of two times the local fluctuation's standard deviation $S_{y'_s}$ of the $y'_i$ values within this fluctuation (Fig. 3, right). In a similar way we define the bandwidths for the s = 1, ... , $\hat{M}$ fluctuations of modeled series *ŷ'* based on $\bar{\hat{y}}'_s$ and $S_{\hat{y}'_s}$. By this, two sequences ($A'_s$) and ($\hat{A}_s$) of local bandwidths are achieved as a structural representation of the single point data ($y'_i$) and ($\hat{y}_i$).

Our basic assumption thus is that the local bandwidths primarily represent the local noise intensities of the series, i.e. the occurrence of local maxima and minima is mainly driven by the noise. Therefore, it is crucial that the arrangement of the sequence boundaries (defined by the local extremes) mainly occurs due to the high frequency variations and thus the local extremes of the noise. As real world signals also contain high frequency variations, the arrangement will in reality possibly be affected by the high frequency variations of the signal, this depending on the spectral properties of the noiseless signal. If the spectrum of the noiseless signal comes close to that of white noise, the interactions between the high frequency variations of signal and noise will distort the arrangement of sequences in a way that the local sequences in average will not reflect the noise intensities. For such datasets SINOMA will not work. However, for signals with spectra different from white noise, the sequence's standard deviations will reflect the noise intensities, but this depending on the departure of the respective spectrum from a white noise spectrum. This is in line with other authors who stated that it is theoretically possible to estimate the EVM parameters if the noiseless data have serial correlations that differ from the coincidential serial correlations of white noise (e.g. Mehra, 1976; Söderström, 1980, 2007). As a consequence, it becomes necessary to determine how 'white' the spectrum of the signal is to gather information on the suitability of the dataset for SINOMA. For this purpose, ad hoc tests based on filtered data have been developed, which will be presented soon (but see also section 6.2).

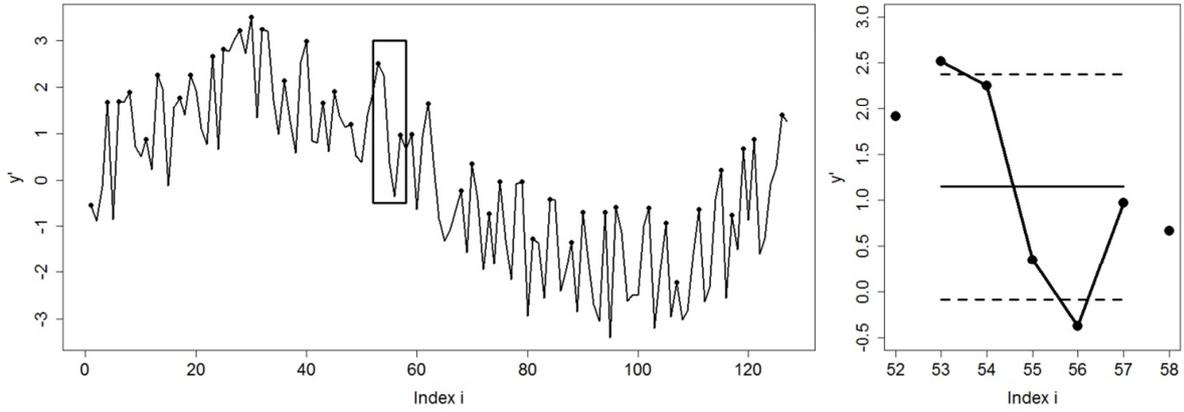

**Fig. 2** Left: Adjustment of the elementary fluctuations, here following the local maxima (dots). Right (margins correspond with the black rectangle in the left image): An arbitrary fluctuation, here consisting of $N_s = 5$ points ($i = 53, ..., 57$), has been enlarged. The solid horizontal line indicates $\overline{y}'_s$, the dashed lines indicate the upper and lower boundaries of $A'_s$ defined as $\overline{y}'_s \pm S_{y'_s}$.

Due to the independent action of noise, both time series $y'$ and $\hat{y}$ will generally not express synchronously arranged local maxima nor minima (i.e. they will not oscillate 'in phase'). Therefore, the comparison of all fluctuation bandwidths $A'_s$ and $\hat{A}_S$ will be carried out on a finer partition, defined through the joint intersections of all $y'$ and $\hat{y}'$ fluctuations (Fig. 3). If, for instance, the first fluctuation $A'_1$ of the $y'$ data is derived from six values and the first two fluctuations $\hat{A}_1$, $\hat{A}_2$ of $\hat{y}'$ are derived from three values each, $A'_1$ has to be compared both to $\hat{A}_1$ and $\hat{A}_2$. In this way we obtain a partition of the discrete time axis $i = 1, ..., N$ into a larger number of $M$ intervals tied together. These are the smallest ones, where either the left or the right end point belongs at least to a local maximum in one of the time series.

We define the local overlap of the time series as the length $O_s \equiv O(A'_s, \hat{A}_s)$ of the intersections of the range intervals of both contributing fluctuations. The definitions so far can be summarized as

$$I'_s := [\overline{y}'_s - S_{y'_s}, \overline{y}'_s + S_{y'_s}] \quad, \quad s = 1, ..., M'$$

$$\hat{I}'_s := [\overline{\hat{y}}'_s - S_{\hat{y}'_s}, \overline{\hat{y}}'_s + S_{\hat{y}'_s}] \quad, \quad s = 1, ..., \hat{M}$$

$$A'_s := 2 \cdot S_{y'_s} \quad, \quad \hat{A}_s := 2 \cdot S_{\hat{y}'_s} \quad, \quad s = 1, ..., M \tag{15a}$$

$$O_s := |I'_s \cap \hat{I}'_s| = \max\left(\min(\overline{y}'_s + S_{y'_s}, \overline{\hat{y}}'_s + S_{\hat{y}'_s}) - \max(\overline{y}'_s - S_{y'_s}, \overline{\hat{y}}'_s - S_{\hat{y}'_s}),\ 0\right) \tag{15b}$$

where, by definition, $0 \leq O_s \leq A'_s, A_s$ holds (with a corresponding relabeling of s according to *M*).

Expressions (15) have to be calculated for each subset of the joint intersection of all fluctuations (indexed by $s = 1, ..., M$). Figure 3 visualizes the respective time series' bandwidth variability along the fluctuations as two 'bands' around the local mean with an 'amplitude' reflecting the local variability of each series as well as the overlap of the respective bandwidths over time.

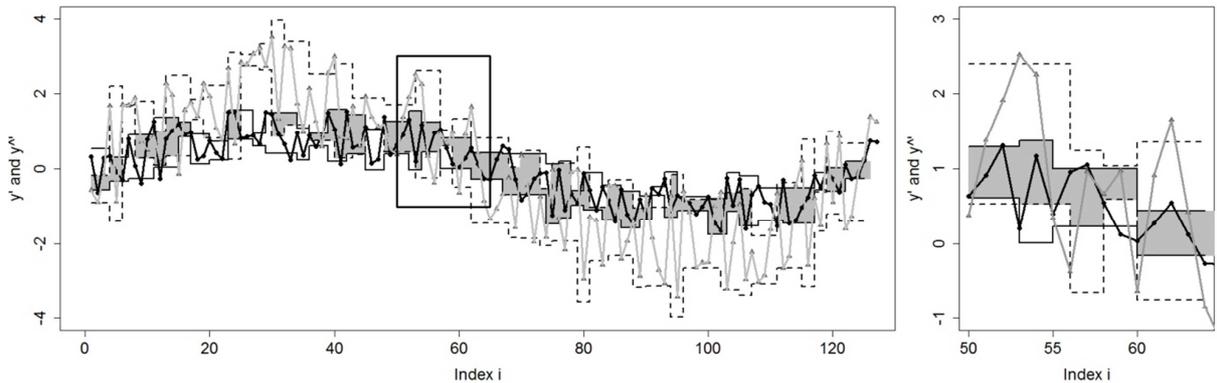

**Fig. 3** Left: Fluctuation bandwidths $A'_s$ (solid line) and $\hat{A}_s$ (dashed line) as 'bands' along the respective time series *y'* (grey points) and *ŷ* (black points). The area shaded in grey indicates the overlapping areas $O_s$ of the respective fluctuations. The black rectangle indicates the margins of the right image which shows a subsequence of the fluctuation bandwidths at a higher resolution

## 3.3 Matching sequential structures

A reasonable measure for the congruence between measured and predicted time series theoretically should be obtained by a comparison of the overlap $O_s$ to the arithmetic mean of the bandwidths $(A_s' + \hat{A}_s)/2$. On the other hand a comparison of $O_s$ to either $A_s'$ or $\hat{A}_s$ indicates to which extent the overlap is explained by either $A_s'$ or $\hat{A}_s$. Therefore, for given time series $y'$ and $\hat{y}'$, we define the local 'explanatory power' $EP_s$ as well as the 'partial explanatory power' indices $EP_s'$ and $\widehat{EP}_s$ as

$$EP_s \equiv EP_s(y', \hat{y}') := \frac{O_s}{(\hat{A}_s + A_s')/2}, \quad EP_s' := \frac{O_s}{A_s'} \quad \widehat{EP}_s := \frac{O_s}{\hat{A}_s} .$$

(16)

If, under 'ideal' conditions (i.e. ignoring unavoidable, but small random sample deviations), the local means $\bar{y}_s'$ and $\bar{\hat{y}}_s'$ and the local bandwidths $A_s'$ and $\hat{A}_s$ are equal and, therefore, $I_s'$ and $\hat{I}_s'$ coincide, the explanatory power $EP_s$ (as well as $EP_s'$ and $\widehat{EP}_s$) reaches the maximal value 1 (i.e. 100% of local variability is mutually explained). In general the respective intervals $I_s'$ and $\hat{I}_s'$ will overlap only partly (or not at all), hence all explanatory power values satisfy

$$0 \leq EP_s \leq 1 , \quad 0 \leq EP_s' \leq 1 , \quad 0 \leq \widehat{EP}_s \leq 1 .$$

(17)

The values of bandwidths $A_s'$ and $\hat{A}_s$ as well as the overlap $O_s$ vary along the sequence of elementary fluctuations, giving rise to a sequence $(EP_s)_{s=1}^M$ of varying values of local explanatory power. Thus it seems to be meaningful to define the arithmetic averages along the sequence,

$$\overline{EP} \equiv \overline{EP}(y', \hat{y}') := \frac{1}{M} \cdot \sum_{s=1}^{M} EP_s(y', \hat{y}') \quad , \tag{18}$$

as the "overall explanatory power" of the reconstruction and in a similar manner for the partial explanatory powers $\overline{\widehat{EP}}$ and $\overline{EP'}$. Although $\overline{EP}$ is not used in the algorithm it can be used to separate the spectral characteristics of the noiseless $x$ and $y$ variables from white noise as discussed in section 6.2. Specific characteristics of $\overline{\widehat{EP}}$ and $\overline{EP'}$ are used to estimate $\lambda$ between $\hat{y}$ and $y'$ which will be explained in detail in the following chapter.

## 4. Sequential Iterative NOise MAtching - SINOMA

### 4.1 Behavior of $\overline{EP'}$ and $\overline{\widehat{EP}}$ over the possible EVM range of $\lambda$

If analyzing the behavior of $\overline{EP'}$ and $\overline{\widehat{EP}}$ over the possible EVM range of $\lambda$ in (7a) from infinity to zero (i.e. varying $\hat{c}$ from $\hat{c}_{OLS}$ to $\hat{c}_{INV}$), $\overline{\widehat{EP}}$ will show its maximum either at or in proximity to $\hat{c}_{OLS}$, while $\overline{EP'}$ shows its maximum either at or in proximity to $\hat{c}_{INV}$. This is because for $\hat{c}_{OLS}$ the fluctuations of $\hat{y}$ data will mostly be encompassed by the fluctuations of y'. By increasing the slope towards $\hat{c}_{INV}$, the fluctuations of $\hat{y}$ will successively enlarge and finally reach a point at which they become larger than those of y'. At $\hat{c}_{INV}$ y' fluctuations will thus mostly be encompassed by $\hat{y}$ (see Fig. 4 for examples). Therefore we introduce the following terms:

$$\overline{EP'}_{INV} = \overline{EP'}(\hat{c}_{INV}); \overline{\widehat{EP}}_{OLS} = \overline{\widehat{EP}}(\hat{c}_{OLS}) \tag{nn}$$

as characteristic values needed for further model identification.

Depending on the error noise in the data, $\overline{\widehat{EP}}_{OLS}$ will either be higher (for OLS conditions), equal to (for RMA conditions), or lower (for INV conditions) than $\overline{EP'}_{INV}$. This is because the error noise in the data affects their local standard deviations and thus $A'_s$ and $\hat{A}_s$. For a lower noise on x (tendency towards OLS conditions), this means that regarding the behavior of $\overline{EP'}$ and $\overline{\widehat{EP}}$ over a range from $\hat{c}_{OLS}$ to $\hat{c}_{INV}$, $\hat{y}$ fluctuations will still be mainly encompassed by *y'* fluctuations for $\hat{c}_{OLS}$. In contrast for $\hat{c}_{INV}$ *y'* fluctuations will only marginally be encompassed by $\hat{y}$ fluctuations. This is because the latter show much lower local amplitudes due to the lower noise on *x* (see Fig. 4, upper panels). For a lower noise on *y* (tendency towards INV conditions), this means that $\hat{y}$ fluctuations will only marginally be encompassed by *y'* fluctuations for $\hat{c}_{OLS}$ but for $\hat{c}_{INV}$ y' fluctuations will completely be encompassed by $\hat{y}$ fluctuations. In this case, the latter show much higher amplitudes due to the higher noise on *x* (see Fig. 4, lower panels). Finally, if the noises of x and y match RMA conditions, $\overline{\widehat{EP}}_{OLS}$ and $\overline{EP'}_{INV}$ should theoretically be identical, although in practice small differences will occur due to finite sample effects (Fig. 4 central panels). To visualize this Fig. 5 shows the behavior of $\overline{EP'}$ and $\overline{\widehat{EP}}$ over the range of $\hat{c}_{EVM}$ slopes for the three example datasets. For a more theoretical explanation of this feature we refer to the appendix (AII).

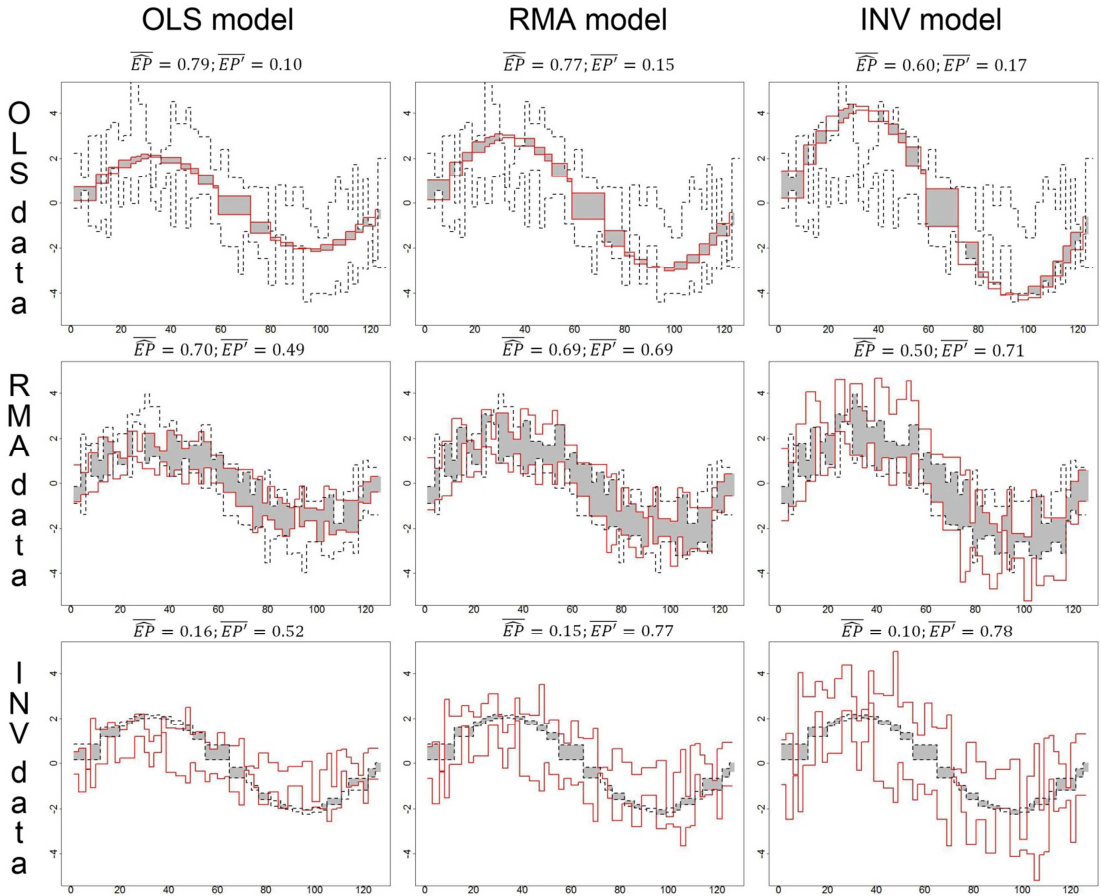

**Fig. 4** Visualization of the local overlaps of the bandwidths $\hat{A}'_s$ and $A'_s$. Each row belongs to a particular data structure (OLS, RMA, INV), each column represents a different model (formulas (9a-c)). The values of $\overline{\widehat{EP}}_{OLS}$ and $\overline{EP'}_{INV}$ are given to demonstrate their dependence on $\hat{c}_{EVM}$ (see also Fig. 5)

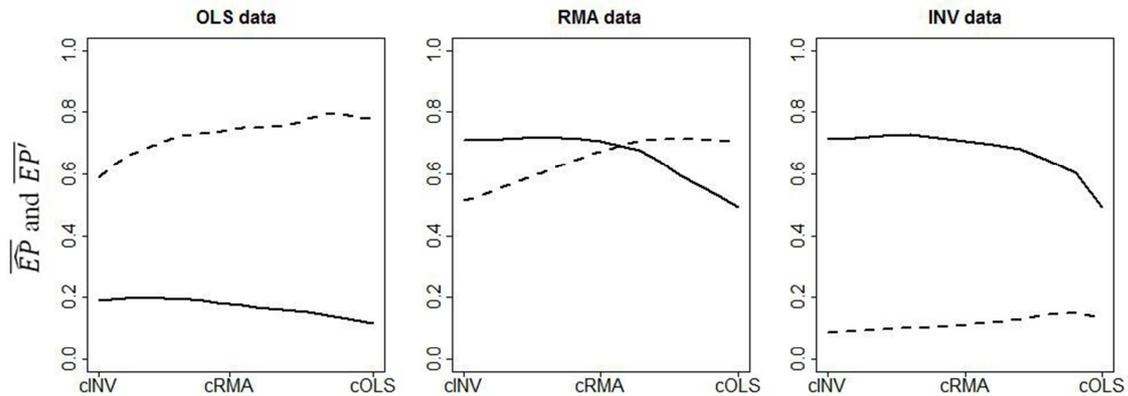

**Fig. 5** $\overline{\widehat{EP}}$ (dashed) and $\overline{EP'}$ (solid) for model data over the range of EVM-model slopes for the three example datasets. For OLS data we have $\overline{\widehat{EP}}_{OLS} > \overline{EP'}_{INV}$, for the RMA data $\overline{\widehat{EP}}_{OLS} \approx \overline{EP'}_{INV}$, and for the INV data $\overline{\widehat{EP}}_{OLS} < \overline{EP'}_{INV}$

## 4.2 $Q_{EP}$ and $\Delta_{EP}$ as criteria for estimating $\lambda$ within SINOMA

Due to the described behavior, we define the ratio between the *independently* maximized mean values,

$$Q_{EP} = \frac{\overline{\widehat{EP}_{OLS}}}{\overline{EP'_{INV}}} \qquad (19)$$

as an indicator of the size of the unknown noise ratio $\lambda$ relative to RMA conditions with

$Q_{EP} > 1$ indicating a lower signal to noise ratio of the y series (i.e. more noise on y),

$Q_{EP} \approx 1$ indicating (close to) RMA conditions, and

$Q_{EP} < 1$ indicating a lower signal to noise ratio of the x series (more noise on x).

$Q_{EP}$ can be used to estimate the noise ratio through formula (A13, see appendix AIII), which then allows for an estimate of the true slope by (7a). A rough estimate of ĉ is provided by

$$\tilde{\hat{c}}_Q = \hat{c}_{INV} \cdot \frac{1-Q_{EP}}{2} + \sqrt{\hat{c}_{RMA}^2 \cdot Q_{EP}^2 + \hat{c}_{INV}^2 \cdot \frac{(1-Q_{EP}^2)^2}{4}} \qquad (20)$$

Another measure of the prevalence of RMA conditions is the difference

$$\Delta_{EP} = (\overline{EP'_{INV}}) - (\overline{\widehat{EP}_{OLS}}) \qquad , \qquad (21)$$

with $\Delta_{EP} < 0$ indicating a lower signal to noise ratio of the y series (more noise on y),

$\Delta_{EP} \approx 0$ indicating (close to) RMA conditions, and

$\Delta_{EP} > 0$ indicating a lower signal to noise ratio of the x series (more noise on x),

with the range $-1 < \Delta_{EP} < 1$.

The sign and value of $\Delta_{EP}$ allows for a slope estimate ĉ for small $\Delta_{EP}$ (i.e. close to RMA conditions) by a simple linear interpolation:

$$\tilde{\hat{c}}_\Delta = \begin{cases} \hat{c}_{RMA} + (\hat{c}_{INV} - \hat{c}_{RMA}) \cdot \Delta_{EP} & if\ \Delta_{EP} \geq 0 \\ \hat{c}_{RMA} + (\hat{c}_{OLS} - \hat{c}_{RMA}) \cdot \Delta_{EP} & if\ \Delta_{EP} < 0 \end{cases} \qquad (22)$$

We want to stress that the values $\tilde{\hat{c}}_Q$ and $\tilde{\hat{c}}_\Delta$ from (20) and (22), respectively, can only be considered as rough estimates for data representing noise conditions in proximity to either OLS or INV data. This is because a $Q_{EP}$ of ∞ or 0 (i.e. $\Delta_{EP}$ of -1 or 1 respectively) cannot be reached as the overlap never is ideal due to random sample deviations. However, towards RMA conditions this error diminishes, wherefore equations (20) and (22) will theoretically deliver precise slope estimates if RMA conditions prevail.

**4.3 Generation of pseudo-RMA conditions within SINOMA**

Therefore, to achieve a best estimate of the true slope, it will be necessary to superimpose the original *x'* or *y'* data with artificial, independent, computer-generated random errors (which ideally meet the general assumptions mentioned in section 2) until RMA conditions are reached, indicated by $Q_{EP} \approx 1$ and/or $\Delta_{EP} \approx 0$. For reasons of simplicity, we only refer to $Q_{EP}$ as a measure of deviation from RMA conditions in the following description of the algorithm. Equivalently, it also is possible to use $\Delta_{EP}$ as such a measure.

SINOMA was specifically developed to iteratively approximate RMA conditions for any kind of initial serial dataset (*x'*, *y'*). In SINOMA, the nature of the error-noise ratio (larger, equal to or smaller than RMA ratio) is estimated by calculating $Q_{EP}$. Depending on the result, either the *x'* data (in case of a stronger noise in the *y'* data, i.e. $Q_{EP} > 1$) or the *y'* data (in case of a stronger noise in the *x'* data, $Q_{EP} < 1$) are superimposed with an independent, artificial "white" noise $\varepsilon_a$ or $\delta_a$ of strong intensity parameterized by its variances. The complementary variable is also superimposed with an artificial noise, but of very weak intensity. This is only done to assure a minimum number of elementary fluctuations in case that the observed variable really was noiseless and only expressed low frequency oscillations (as e.g. the sine waves used as examples). The modified series have the structure

$$x_i'' := x_i + \varepsilon_i + \varepsilon_{a,i} \quad, \quad y_i'' := y_i + \delta_i + \delta_{a,i} \quad, \quad i = 1, ..., N \quad, \tag{23}$$

where all values ($\varepsilon_{a,i}$) and ($\delta_{a,i}$) are mutually independent, identically and uniformly distributed random numbers with zero mean and variances $S_{\varepsilon_a}^2$ and $S_{\delta_a}^2$, respectively.

$Q_{EP}$ is calculated for the new series $x''$ and $y''$. In case the relation of $Q_{EP}$ to 1 (being either > or <) remains unchanged, the added noise was too weak to generate RMA conditions. In this case an even stronger artificial noise is superimposed to the same data set as before (either $x'$ or $y'$). In case the relation of $Q_{EP}$ to 1 changes the superimposed noise was too strong and the RMA point has been passed. Then the noise strength is reduced a little until the relation of $Q_{EP}$ to 1 again changes. Each time when the relation of $Q_{EP}$ to 1 changes (forth and back), a new preliminary slope estimate $\tilde{\tilde{c}}$ can be calculated by equation (20). This procedure is iteratively repeated until $Q_{EP}$ values are obtained being close enough to one (e.g. $|Q_{EP} - 1| < 0.01$). In practice it is however very unlikely that $Q_{EP}$ becomes exactly one because of the unavoidable variations of finite elementary fluctuation samples and, hence, violations of the general assumptions. In our test runs, the value of $Q_{EP}$ finally fluctuated erratically around one. As a break criterion for the iteration, we therefore demanded that the difference of $\tilde{\tilde{c}}$ from (20) to the slope $\hat{c}_{RMA}$ in (9b) becomes less than a certain threshold (e.g. 0.01). At this point we claim that the modified data set has reached pseudo-RMA conditions. Then, the slope estimates (20) can be interpreted as a very small correction to the $\hat{c}_{RMA}$ value of this noise-specific dataset. In Table 4 we present some numerical results derived from the application of the SINOMA algorithm to the example data sets.

From the final slope estimate $\hat{c} = \hat{c}_{EVM} \approx c$ the original noise ratio $\lambda \equiv \lambda_{EVM}(x', y')$ can be calculated according to (8) or, more directly, using (12). Then, by definition of the RMA case, the noise ratio $\lambda''$ of the modified time series ($x''$, $y''$) with mutually independent extra noise contributions $S_{\delta_a}^2$ and $S_{\varepsilon_a}^2$ is calculated as

$$\lambda''_{RMA}(x'',y'') = \frac{S_\delta^2 + S_{\delta_a}^2}{S_\varepsilon^2 + S_{\varepsilon_a}^2} = \frac{\lambda_{EVM} \cdot S_\varepsilon^2 + S_{\delta_a}^2}{S_\varepsilon^2 + S_{\varepsilon_a}^2} \tag{24}$$

and should be equal to $\hat{c}^2_{RMA}$ (cf. (9b)). This equation can be solved for $S_\varepsilon^2$ which finally gives

$$S_\varepsilon^2 = \frac{S_{\delta_a}^2 - \lambda''_{RMA} \cdot S_{\varepsilon_a}^2}{\lambda''_{RMA} - \lambda_{EVM}} \quad , \quad S_\delta^2 = \lambda_{EVM} \cdot S_\varepsilon^2 \tag{25}$$

as the best estimates of the (formerly unknown) noise variances $S_\varepsilon^2$ and $S_\delta^2$ embedded in the original observations $x'$ and $y'$. Given (5), it now becomes possible to estimate the standard deviations $S_x$ and $S_y$ of the noiseless data.

Table 4: Numerical values of $Q_{EP}$ and $\tilde{c}$ according to (20) and (21) after the first step of the algorithm and after 50 iterations, generating nearly RMA conditions for the three (iteratively modified) example datasets. The first slope estimate $\tilde{c}$ based on (20) is too high (or too low) for OLS (or INV) data, while the slope estimate for the RMA example already is very close to the true slope. After 50 iterations of the algorithm $\tilde{c}$ estimates nearly the true value of 2.1 in all three cases.

| dataset | first estimate of $Q_{EP}$ | first estimate of $\tilde{c}$ from (20) | SINOMA estimate of $\tilde{c}$ |
|---|---|---|---|
| OLS data | 2.948 | 2.14 | 2.06 |
| RMA data | 1.001 | 2.09 | 2.09 |
| INV data | 0.578 | 1.88 | 2.15 |

## 5. Validation of SINOMA

To test the validity of SINOMA on a data set that better represents real time series, it was applied to a pseudo-proxy dataset. Here, the noisy $x'$ and $y'$ data were respectively generated by adding two different artificial noises to a temperature series obtained from a climate simulation with an atmosphere-ocean general circulation model covering the past 1000 years (von Storch et al., 2004). These data reflect the simulation of annual mean temperatures of one grid cell representative for the subpolar region of Northern Europe in the period 1890-1990 (mean: 1.232 °C, sd: 1.092 °C; see Fig. 8) and were subsequently normalized to a standard deviation of one. The relationship between (noiseless) $x$ and $y$ was simply put as $y = 1 \cdot x$ (i.e. $c = 1$, $c_0 = 0$). The noise intensities were thereby selected in a range that the coefficient of determination $R^2$ of classical OLS calculations was approximately 0.3. This $R^2$ value is comparable to what frequently is achieved in terms of climatological reconstructions from tree-rings (e.g. Fritts, 1976).

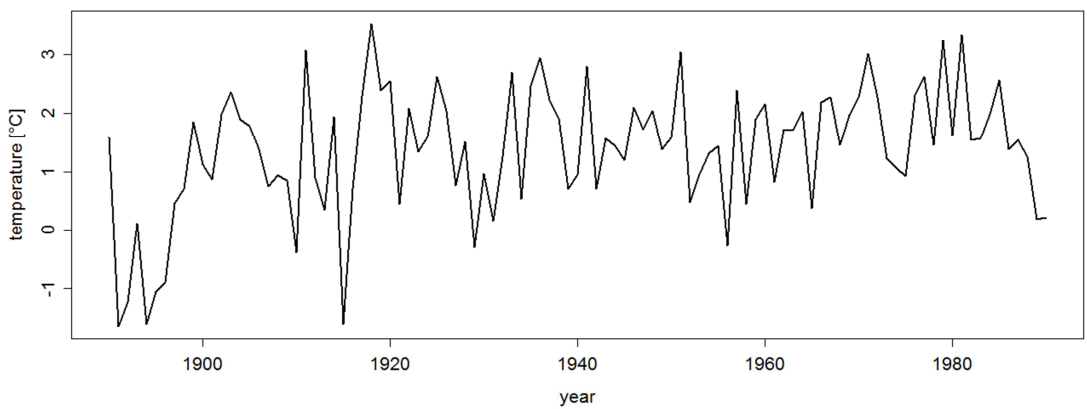

**Fig. 6** Annual mean near-surface air temperature simulated in coupled atmosphere-ocean simulation with the climate model ECHO-G in a model grid-cell located in the Artic region. The model was driven by variable realistic external climate forcing (von Storch et al., 2004)

To reflect the possible variety of the error noises, ten different pseudo-proxy data sets with different noise ratios were generated (cf. Table 5). The EVM slope estimates through formula (7a) from these pseudo-proxy datasets were always close to the true slope and, of course, better than the estimates simply achieved by the OLS, RMA and INV formula, respectively. They were, however, never exactly equal to 1, because unavoidable small 'errors' were caused by the violation of the general assumptions through covariances of the superimposed noise data (covariances in the range of ±0.001 to ±0.1). But these noise covariances can also be taken as more representative for realistic conditions, as the 'noise' in many cases is, to some extent, either correlated with one of the variables and/or the respective other noise. SINOMA was then applied ten times (with 40 iterations each) to an arbitrary, randomly selected pseudo-proxy dataset, namely E1_Test4 (see Tables 5 and 6). A ten-fold replication of the algorithm was carried out, because in each iterative step the unavoidable violations of the ideal assumptions on the noise terms (being independent, centered, no covariance etc.) caused small fluctuations of the numerical slope estimates around the optimal estimate. However, on average SINOMA achieved very accurate slope estimates (close to the true value 1) which were always much better than the estimates derived from simple OLS, RMA or INV models. While the sinus-like data in chapter 4 represent a rather smooth signal disturbed by noise, the latter example shows that also seemingly "irregular" time series like the ERIK data can be successfully processed with SINOMA. The accuracy is as good as the empirical slope estimates based on known noise ratio and formula (7a), but our algorithm works *without* any information about the noise component in the observed variables. Besides obtaining the slope estimate, SINOMA allows for estimating the standard deviations of the noiseless $x$ and $y$ data and the respective error noises by applying (5) and (25) (see Table 6). A comparison of tables 5 and 6 shows that even for this rather irregular dataset, the average $\lambda_{EVM}$ of 0.229 obtained by ten calculations is in good proximity to the true error noise ratio of 0.182. Also the standard deviations of the noiseless $x$ and $y$ data are well estimated (1.144 and 1.069, respectively, compared to 1.092).

Table 5: Data set ERIK_grid 1 ($N = 101$) from von Storch et al., (2004) superimposed with computer-generated random errors with *predetermined* noise ratio $\lambda$. The slope $\hat{c}_{EVM}(\lambda)$ was calculated by equation (7a). For computer-generated random noise realizations the empirical range of covariances among each other and to the variables was typically ±0.001 to ±0.1.

| data set | $\lambda$ | $R^2$ | true $c$ | $\hat{c}_{OLS}$ | $\hat{c}_{RMA}$ | $\hat{c}_{INV}$ | $\hat{c}_{EVM}(\lambda)$ |
|---|---|---|---|---|---|---|---|
| E1_Test1 | 0.096 | 0.314 | 1.0 | 0.331 | 0.591 | 1.053 | 0.873 |
| E1_Test2 | 0.118 | 0.275 | 1.0 | 0.364 | 0.695 | 1.326 | 1.108 |
| E1_Test3 | 0.142 | 0.335 | 1.0 | 0.458 | 0.791 | 1.366 | 1.176 |
| E1_Test4 | 0.182 | 0.295 | 1.0 | 0.394 | 0.726 | 1.335 | 1.048 |
| E1_Test5 | 0.715 | 0.277 | 1.0 | 0.486 | 0.922 | 1.752 | 0.997 |
| E1_Test6 | 1.481 | 0.271 | 1.0 | 0.558 | 1.071 | 2.057 | 0.953 |
| E1_Test7 | 2.972 | 0.285 | 1.0 | 0.733 | 1.375 | 2.577 | 1.138 |
| E1_Test8 | 9.790 | 0.260 | 1.0 | 0.758 | 1.488 | 2.921 | 0.899 |
| E1_Test9 | 10.500 | 0.337 | 1.0 | 0.931 | 1.603 | 2.762 | 1.092 |
| E1_Test10 | 16.700 | 0.324 | 1.0 | 0.955 | 1.679 | 2.951 | 1.070 |
| mean $\bar{\hat{c}}$ | | | | | | | 1.035 |
| st. dev. $S_{\hat{c}}$ | | | | | | | 0.102 |

Table 6: Test data set E1_Test4 (from Table 5), ten times superimposed with additional computer-generated noise of different intensities (denoted by suffixes AN1–10). The slope estimate $\hat{c}_{SNM}$ was calculated by the SINOMA algorithm proposed in chapter 4. The standard deviations of the noiseless $x'$ and $y'$ data and the respective error noises were achieved by (5) and (30).

| data set | $\hat{c}_{SNM}$ | $s_{\varepsilon_a}$ | $s_{\delta_a}$ | $\lambda_{EVM}$ | $s_\varepsilon$ | $s_\delta$ | $s_x$ | $s_y$ |
|---|---|---|---|---|---|---|---|---|
| E1_Test4+AN1 | 1.004 | 0.028 | 1.112 | 0.215 | 1.248 | 0.579 | 1.127 | 1.074 |
| E1_Test4+AN2 | 0.956 | 0.028 | 0.951 | 0.255 | 1.171 | 0.591 | 1.207 | 1.068 |
| E1_Test4+AN3 | 0.976 | 0.028 | 1.031 | 0.237 | 1.218 | 0.593 | 1.160 | 1.067 |
| E1_Test4+AN4 | 1.022 | 0.028 | 1.110 | 0.214 | 1.185 | 0.548 | 1.193 | 1.090 |
| E1_Test4+AN5 | 1.023 | 0.028 | 1.182 | 0.200 | 1.285 | 0.575 | 1.085 | 1.077 |
| E1_Test4+AN6 | 0.996 | 0.028 | 1.146 | 0.221 | 1.305 | 0.613 | 1.061 | 1.055 |
| E1_Test4+AN7 | 0.987 | 0.028 | 1.057 | 0.229 | 1.224 | 0.586 | 1.153 | 1.071 |
| E1_Test4+AN8 | 0.930 | 0.028 | 0.924 | 0.277 | 1.204 | 0.634 | 1.174 | 1.043 |
| E1_Test4+AN9 | 0.959 | 0.028 | 0.989 | 0.252 | 1.210 | 0.608 | 1.167 | 1.058 |
| E1_Test4+AN10 | 1.036 | 0.028 | 1.189 | 0.190 | 1.265 | 0.551 | 1.108 | 1.089 |
| mean | 0.990 | | | 0.229 | 1.232 | 0.588 | 1.144 | 1.069 |
| st. dev. | 0.034 | | | 0.027 | 0.043 | 0.027 | 0.047 | 0.015 |

## 6. Discussion and Conclusion

### 6.1 The potential of SINOMA

In this paper, we have introduced a new approach (SINOMA) to derive an unbiased estimation of the true slope of a linear relationship between serially realized, noisy *x* and *y* data without any extraneous knowledge on the noise variance ratio of these variables. SINOMA is able to estimate the noise variances of both noisy variables and the respective noise variance ratio directly from the two noisy serial data streams. This is possible because the two serial data streams have a linear relation and thus a common sequential structure allowing the separation of signal and noise. It is essential for our novel approach that, in contrast to standard regression procedures such as OLS, RMA and INV models, the estimation of the model slope is not only based on the simple comparison of single pairs (*x'*, *y'*) of noisy measurements. Additionally, local 'bandwidths' of 'elementary fluctuations' in the sequential realizations of the $y'_i$ and $\hat{y}'_i$ data are compared, thereby evaluating the structural similarity between the two serial data streams. The 'best' slope estimate $\hat{c}$ is iteratively determined by I) artificially generating RMA conditions (validated by $Q\overline{EP}_{max} \approx 1$) through adding white noise and II) extracting the slope estimate at this special point by either the $c_{RMA}$ slope estimate which is appropriate under these conditions or application of the error noise estimates to (20).

To our knowledge, the approach of SINOMA is completely new, as noisy (x', y') data so far could only be handled correctly if either the error variance of the predictor variable or the error variance of the predictand variable or the error noise ratio of both was known, or at least good estimates of these noise measures were available (e.g. Kutzbach et al., 2011; Moberg and Brattström, 2011). However, in most cases, there is no extraneous knowledge about the noise of the x and y data available. Hegerl et al. (2007) and Ammann et al. (2010) proposed other possibilities for empirically estimating the error variances directly from two noisy time series in the context of paleo-climate reconstruction. Hegerl et al. (2007) estimate the proxy (predictor) error variance by varying it over a wide range and selecting the error variance that produced the lowest deviations between reconstructed and observed climatic time series (evaluated by an F-test). In our view, this approach should lead to a systematic preference for low proxy error variances since the model evaluation is again based on the comparison of single pairs (x', y') of noisy measurements and noisy reconstructed values. However, the residuals between the noisy calibration and reconstructed data pairs should be minimized per definition by OLS regression (which will on the other hand deliver attenuated slope estimates). The approach of Ammann et al. (2010), who proposed estimating the proxy error variance by minimizing the bias between predicted (reconstructed) and independent validation datasets, (by cross-validation) appears to us more reasonable. However, the authors did not express how the validation bias should be evaluated, which is in our view the central question (see discussion above). As statistical measures that directly compare single pairs (x', y') of noisy measurements are inappropriate, lowpass filtering would be required; however, then the question arises as to whether the calibration time series of the instrumentally measured climatic variable is long enough (has enough degrees of freedom for a fivefold cross-validation as suggested by Ammann et al. (2010).

As our approach is able to estimate the error variances only from the two noisy data streams, we are convinced that the proposed approach will substantially contribute to many research fields dealing with the identification of linear relationships between pairs of serially realized data, such as paleo-climatology, dendrochronology, and remote sensing. With the examples given in this manuscript, we were able to show that SINOMA is able to estimate the true slope of a linear relationship between noisy data with a high accuracy. The estimates are numerically equal to those obtained from OLS, RMA and INV models if the error-related requirements of these model approaches are fulfilled. They are (within random deviations) much better than those other models in the cases in which their underlying assumptions are not fulfilled.

## 6.2 Constraints of SINOMA

Even though SINOMA estimates model slopes with a high accuracy, we want to stress that for all but true OLS conditions (noiseless $x$ data) the modeled data $\hat{y}'$ will still be erroneous – in the sense that they will be disturbed by noise. This is because the modeled data (apart from exact OLS conditions) have to be calculated from noisy $x'$ data. In case of RMA conditions (variance matching) the comparison of modeled and measured $y$ data will give the impression that the model output represents the measured series fairly well since the variance of the (noisy) calibration data and the data reconstructed from the noisy predictor data using the RMA model are by definition equal. However, this is only true for the *noisy* measured $y'$ data, while they do not directly represent the *noiseless* $y$ data (see also Fig 1 lower central panel). The larger the noise in the $x'$ data, the larger the error variance of the modeled (predicted, reconstructed) data becomes – despite a high accuracy of the slope estimate. It is important to note that the "inflated" high-frequency noise variance of the predicted data of errors-in-variables approaches compared to OLS (and probably often also RMA) approaches (National Research Council, 2006; Tingley and Li, 2012) should not be seen as a problem because it is unavoidable when the correct model (in terms of unbiased parameters) is chosen and applied to noisy predictor (proxy) data (see also Christiansen 2012). This statement implies that the suitability of traditional measures for model performance RMSE, $r^2$, RE and CE (see National Research Council (2006) for their definitions) is not a general property but rather depends on the particular goal of the analysis.

An approach to solve this problem is to 'smooth' the data by application of lowpass filters. Under white-noise conditions, in particular the high-frequency oscillations of the serial data are affected by noise. Therefore, lowpass filtering of the data will allow for a comparison of the $y'$ and $\hat{y}'$ data with a weaker influence of the white noise on this comparison. This will however decrease the temporal (or spatial, depending on the nature of the serial data streams) resolution of the filtered model data, i.e. the specific change after one time-step will be blurred due to the filtering. But in contrast, the low-frequency oscillations of the target variable will be represented with a high accuracy regarding its absolute, noiseless amplitudes, thus allowing for more precise reconstructions. For instance, if climate variables are reconstructed from noisy proxies at an annual resolution, it will not be possible to calculate the absolute values with a high, annual resolution. In contrast the low-frequency climate fluctuations – which anyway are more interesting in terms of climate reconstructions – will be reliably reconstructed.

It is important to stress that SINOMA will only deliver reliable slope estimates if the arrangement of elementary fluctuations occurs due to the high-frequency structures present in the noise. This will only occur if the noiseless signal has different spectral characteristics than white noise. Due to this, it becomes necessary to apply ad-hoc tests to the data before using SINOMA. These tests should evaluate whether the spectral properties of the data differ from white noise. Therefore, parallel to the development of SINOMA, we have developed an ad-hoc test which is based on the overall explanatory power ($\overline{EP}$) calculated for filtered data along a gradient of different filter lengths. This procedure allows to I) separate white noise processes from other processes and II) to identify filter lengths which are suited best to express possible low frequency structures of the signal behind the noise. By this it then also becomes possible to choose adequate filters in order to acquire noiseless low-frequency reconstructions of the target variable as claimed in the previous paragraph.

Another constraint of SINOMA is that a certain minimum of fluctuations (each for $y'$ and $\hat{y}'$ data) is required to provide a reliable slope estimate. This is because the white noise assumption will be violated to a certain degree for all of the elementary fluctuations due to the low sample size within each fluctuation. In theory however, these violations will be evened out by averaging over several fluctuations. As for consistent estimators, the precision of the estimate increased with sample size. Through numerous test runs with a variety of data we came to the conclusion that six elementary fluctuations are the minimum requirement for successful slope estimates.

Another prerequisite of SINOMA is that the true slope $\hat{c}$ must be in proximity to one, otherwise the contribution of a low error noise variance could be overwhelmed by the steep deterministic change of the non-noisy variables $x$ and $y$ even within one elementary fluctuation. This problem (characterized by $\hat{c}_{\text{OLS}} \gg 1$) can be avoided if the $y'$ data are divided by $\hat{c}_{\text{OLS}}$ lowering the slope in a way to let the error noise related fluctuations become dominant. For convenience, SINOMA assumes that the slope relating both time series is positive. If the $x'$ and $y'$ data were negatively correlated, a simple change of the sign of $y'$ series will make the slope positive.

Further critical issues are related to the general EVM assumptions (see also section 2). Since SINOMA works with noisy serial data, the noise terms of the predictor and predictand variables are likely to be serially correlated with themselves and/or with the predictor and predictand variables. Thus, the assumption of independence of the errors will often be violated in reality (Robinson 1986; Christiansen et al., 2009; Moberg and Brattström, 2011). It will therefore be important to investigate more in detail the behavior of SINOMA under violated error assumptions. Buras et al. (submitted) have undertaken a first approach to investigate the behavior of SINOMA for 4400 pseudo-proxy datasets which have correlations of differing strengths between noise and signal (i.e. 'red noise'). They could show that all regression models (OLS, RMA, INV, EVM and SINOMA) were affected by red noise, however the strength of effect did not differ significantly between EVM and SINOMA. This indicates that even under red noise conditions SINOMA is competitive with the theoretically most precise approach (EVM).

In conclusion, SINOMA will work reliably as indicated by several numerical tests (here and by Buras et al., submitted) as long as the described requirements are fulfilled. These are:

- a signal with spectral characteristics different from white noise,
- at least six elementary fluctuations,
- low covariances between the variables and their noise errors,
- a positive, not too steep slope.

**6.3 Outlook**

As next important steps, SINOMA should be evaluated against other recently proposed errors-in-variables methods using pseudoproxy approaches and "real-world" tests. Of highest interest in the scope of the paleoclimate reconstruction debate would be a thorough comparison of SINOMA with the methodologies outlined by Ammann et al. (2010), Christiansen (2010), Tingley et al. (2010a, b) and Moberg and Brattström (2011), respectively. Special emphasis should be put on the effects of serially auto- and co-correlated errors and empirical analysis of naturally occurring error properties of frequently studied serial noisy data streams. A first approach into this direction has been undertaken by Buras et al. (submitted). Since SINOMA is able to empirically estimate the noise variances of two noisy measured variables, it offers new interesting opportunities to study these noise processes, e.g. by comparison with independent (extraneous) noise estimates like replicate measurements or ensemble variability analyses (e.g. Kutzbach et al., 2011; Moberg and Brattström, 2011). A user-friendly program and a package for the statistical programming language 'R' will be available soon, in order to supply the scientific community with this new modeling tool, this allowing for an open discussion within the palaeo-climate community (amongst others).

## 6.4 Conclusions

In conclusion, this manuscript introduces a new approach (SINOMA) to model linear relationships between two noisy variables, realized as serial data. The examples provided reveal that SINOMA is able to estimate the slope between the variables of interest with a high accuracy and much better than the commonly applied OLS, INV or RMA models. Further, SINOMA allows for estimating the standard deviations of the noiseless variables as well as the error noise ratio of the data. The application of SINOMA to already published climate reconstructions based on serially realized noisy data will likely result in different amplitudes of those reconstructions. SINOMA thus has the potential to reframe our knowledge that is based upon those climate reconstructions.

## Acknowledgements

This study was conducted within the project „Neue statistische Verfahren für amplitudenoptimierte Paläoklimarekonstruktionen aus zeitlich hochaufgelösten, verrauschten Proxydaten" which is funded by the German Research Foundation (KU 1418/5-1, WI 2680/5-1). Further, the study is a contribution to the Virtual Institute of Integrated Climate and Landscape Evolution Analysis –ICLEA– of the Helmholtz Association. L. Kutzbach was supported through the Cluster of Excellence "CliSAP" (EXC177), University of Hamburg, funded through the DFG.

**References**

Agüero JC and Goodwin GC (2008) Identifiability of errors in variables dynamic systems. Automatica, 44(2): 371-382.

Akritas MG, and Bershady, MA (1996) Linear regression for astronomical data with measurement errors and intrinsic scatter. arXiv preprint astro-ph/9605002. Astrophysical Journal, 470: 706.

Allen MR, Stott PA (2003) Estimating signal amplitudes in optimal fingerprinting. Part I: Theory. Climate Dyn., 21: 477–491.

Ammann CM, Genton MG, and Li B (2010) Technical Note: Correcting for signal attenuation from noisy proxy data in climate reconstructions. Clim. Past, 6, 273-279.

Anderson, B. D. O., & Deistler, M. (1984) Identifiability in dynamic errors-in-variables model. Journal of Time Series Analysis, 5(1): 1–13.

Brohan P, Kennedy JJ, Harris I, Tett SFB, and Jones PD (2006) Uncertainty estimates in regional and global observed temperature changes: A new data set from 1850. Journal of Geophysical Research, 2: 99–113.

Buras A, Thees B, Czymzik M, Dräger N, Heine I, Kienel U, Neugebauer I, Ott F, Scharnweber T, Simard S, Słowińska S, Słowiński M, Tecklenburg C, Zawiska I, Kutzbach L, Brauer A, and Wilmking M (submitted) SINOMA – a better tool for proxy based reconstructions?

Carpenter SR, Cottingham KL, and Stow CA (1994) Fitting predator-prey models to time series with observation errors. Ecology, 75: 1254-1264.

Castaldi, P., & Soverini, U. (1996) Identification of dynamic errors-in-variables models. Automatica, 32(4): 631-636.

Cheng CL and Van Ness JW (1999) Statistical regression with measurement error (Vol. 6). London: Arnold.

Christiansen B, Schmith T, and Thejll P (2009) A surrogate ensemble study of climate reconstruction methods: Stochasticity and robustness. J Clim, 22: 951-976.

Christiansen B (2011) Reconstructing the NH Mean Temperature: Can underestimation of trends and variability be avoided? J Clim, 24: 674-692.

Christiansen B and Ljungqvist FC (2011) Reconstruction of the extra- tropical NH mean temperature over the last millennium with a method that preserves low-frequency variability. J Clim, 24: 6013–6034.

Christiansen B (2012) Reply to "Comments on 'Reconstructing the NH Mean Temperature: Can Underestimation of Trends and Variability be Avoided?'". J Clim: 25, 3447-3452.

Christiansen B and Ljungqvist FC (2012a) The extra-tropical NH temperature in the last two millennia: Reconstructions of low-frequency variability. Clim. Past, 8: 765-786.

Christiansen B and Ljungqvist FC (2012b) Reply to "Comments on 'Reconstruction of the Extratropical NH Mean Temperature over the Last Millennium with a Method That Preserves Low-Frequency Variability'". J Clim, 25: 7998-8003.

Coelho CAS, Pezzulli S, Balmaseda M, Doblas-Reyes FJ, and Stephenson DB (2004) Forecast calibration and combination: A simple Bayesian approach for ENSO. J Clim, 17: 1504–1516.

Deistler M and Anderson BD (1989) Linear dynamic errors-in-variables models: some structure theory. Journal of Econometrics, 41(1): 39-63.

Esper J, Cook ER, and Schweingruber FH, (2002) Low-frequency signals in long tree-ring chronologies for reconstructing past temperature variability. Science, 295: 2250–2253.

Frisch R (1934) Statistical confluence analysis by means of complete regression systems. Technical Report 5. Univ. of Oslo, Economics Institute, Norway.

Fuller WA (1987) Measurement error models. Wiley Series in Probability and Statistics. Wiley, New York.

Fritts HC (1976): Tree Rings and Climate. Academic Press, New York.

Fritts HC, Guiot J, Gordon GA, and Schweingruber FH (1990) Methods of calibration, verification, and reconstruction. In: Cook ER, Kairiukstis LA (eds) Methods of dendrochronology: Applications in the environmental sciences. Kluwer, Dordrecht, pp 163–217.

Hartung J (1999) Statistik, Lehr- und Handbuch der angewandten Statistik. R. Oldenbourg Verlag München Wien.

Hausman J (2001) Mismeasured Variables in Econometric Analysis Problems from the Right and Problems fromthe Left. The Journal of Economic Perspectives, 15: 57-67.

Hegerl GC, Crowley TJ, Allen M, Hyde WT, Pollack HN, Smerdon J and Zorita E (2007) Detection of human influence on a new, validated 1500-year temperature reconstruction. J Clim, 20: 650–666.

Jones PD, Osborn TJ, Briffa KR (1997) Estimating sampling errors in large-scale temperature averages. J Clim, 10: 2548–2568.

Kendall MG and Stuart A (1973) The Advanced Theory of Statistics, vol. 2. Griffin and Co., London.

Kutzbach L, Thees B, and Wilmking M (2011) Identification of linear relationships from noisy data using errors-in-variables models –relevance for reconstruction of past climate from tree-ring and other proxy information, Climatic Change, 105: 155-177.

Matsumoto M and Nishimura T (1998) Mersenne Twister: a 623-dimensionally equidistributed uniform pseudo-random number generator. ACM Transactions on Modeling and Computer Simulations (TOMACS), 8: 3-30.

McArdle BH (1988) The structural relationship: regression in biology. Can J Zool, 66: 2329–2339.

Mehra RK (1976) Identification and estimation of the error-in-variables model (EVM) in structural form. In Wets RJ-B (ed) Stochastic Systems: Modeling, Identification and Optimization, I. Springer, Berlin, Heidelberg, pp. 191-210.

Moberg A, Sonechkin DM, Holmgren K, Datsenko NM, and Karlén W (2005) Highly variable Northern Hemisphere temperatures reconstructed from low- and high-resolution proxy data. Nature, 3265: 4 pp.

Moberg A and Brattström G (2011) Prediction intervals for climate reconstructions with autocorrelated noise - An analysis of ordinary least squares and measurement error methods. Palaeogeography, Palaeoclimatology, Palaeoecology, 308: 313-329.

Moberg A. (2012) Comments on "Reconstruction of the extra-tropical NH mean temperature over the last millennium with a method that preserves low-frequency variability". J Clim 25: 7991-7997.

National Research Council (2006) Surface Temperature Reconstructions for the Last 2,000 Years. National Academic Press, Washington.

Reiersøl O (1950) Identifiability of a linear relation between variables which are subject to error. Econometrica 18: 375-389.

Riedwyl N, Küttel M, Luterbacher J, and Wanner H (2009) Comparison of climate field reconstruction techniques: Application to Europe. Climate Dynamics, 32: 381-395.

Robinson PM (1986) On the errors-in-variables problem for time series. Journal of multivariate analysis, 19: 240-250.

Sachs HM, Webb T III, Clark DR (1977) Paleoecological transfer functions. Ann Rev Earth Planet Sci, 5: 159-178.

Shumway RH and Stoffer DS, (2011) Time Series Analyses and its applications. 3$^{rd}$ edition, Springer, New York, 606 pp.

Söderström T (1980) Spectral decomposition with application to identification. In: Archetti F and Cugiani M (eds.), Numerical techniques for stochastic systems Amsterdam: North-Holland, pp. 59–79.

Söderström T (2007) Errors-in-variables methods in system identification. Automatica, 43(6): 939-958.

Solo V (1986) Identifiability of time series models with errors in variables. In Gani J and Priestley MB (eds), Essays in time series and allied processes, 23: 63–71.

Stoica P and Nehorai A (1987) On the uniqueness of prediction error models for systems with noisy input–output data. Automatica, 23(4): 541–543.

Thees B, Kutzbach L, Wilmking M, and Zorita E (2009) Ein Bewertungsmaß für die amplitudentreue regressive Abbildung von verrauschten Daten im Rahmen einer iterativen „Errors in Variables"- Modellierung (EVM), GKSS-report 2009/8.

Tingley, MP and Huybers P (2010a) A Bayesian Algorithm for Reconstructing Climate Anomalies in Space and Time. Part 1: Development and applications to paleoclimate reconstruction problems. J Clim, 23: 2759-2781.

Tingley, MP and Huybers P (2010b) A Bayesian Algorithm for Reconstructing Climate Anomalies in Space and Time. Part 2: Comparison with the Regularized Expectation-Maximization Algorithm. J Clim, 23: 2782–2800.

Tingley, MP and Li B (2012) Comments on "Reconstructing the NH mean temperature: Can underestimation of trends and variability be avoided?" J Clim, 25: 3441-3446.

Tingley MP, Craigmile PF, Haran M, Li B, Mannshardt-Shamseldin E, and Rajaratnam B (2012) Piecing together the past: Statistical insights into paleoclimatic reconstructions. Quaternary Science Reviews, 35: 1-22.

von Storch H, Zorita E, Jones J, Dimitriev Y, Trett S, and Gonzales-Rouco F (2004) Reconstructing past climate from noisy data. Science, 306: 679-682.

Zorita E, Gonzales-Rouco F, and Legutke S (2003) Testing the Mann et al. (1998) approach to Paleoclimatic reconstructions in the context of a 1000 yr control simulation with the ECHO-G coupled climate model. Journal of Climate, 16: 1378-1390.

**Appendix**

Table A1: Overview on the nomenclature of variables used within the manuscript. Further explanations are given in the text.

| | |
|---|---|
| $i = 1, ..., N$ | sample index for $x, y, x', y', \hat{y}', ...$ in sequences of $N$ points |
| $x_i, y_i$ | true values of explanatory and response variable of $i$th observation |
| $x'_i, y'_i$ | observed values disturbed by noise |
| $\varepsilon_i, \delta_i$ | random error values ('noise') related to $x$ and $y$ ($x'_i = x_i + \varepsilon_i$, $y'_i = y_i + \delta_i$) |
| $\hat{y}'_i$ | predicted value of response variable estimated from explanatory value $x'$ |
| $S^2_{x'}, S^2_{y'}, S^2_{\varepsilon}, S^2_{\delta}$ | series variance of the noisy $x'$ and $y'$ values and their respective noises |
| $S_{x'y'}$ | series covariance of the noisy $x'$ and $y'$ values |
| $\lambda = S^2_{\delta}/S^2_{\varepsilon}$ | noise ratio (ratio of the variances of the error-noises) |
| $R^2 := \dfrac{S^2_{x'y'}}{S^2_{x'} \cdot S^2_{y'}}$ | coefficient of determination (i.e. explained variance in OLS regression) |
| $\hat{c}, \hat{c}_0$ | estimated values of the true (unknown) intercept slope $c$ and intercept $c_0$ |
| $s$ | Elementary fluctuation index for sequences $y'$ and $\hat{y}'$ |
| $M', \widehat{M}, M$ | number of elementary fluctuations in sequences $y'$ and $\hat{y}'$ and in joint partition of time axis |
| $A'_s \equiv A_s(y')$, $\hat{A}_s \equiv A_s(\hat{y}')$ | local area of equi-amplitude, defined by subsequent local extrema in the $y'$ and $\hat{y}'$ series, respectively |
| $O_s \equiv O(A'_s, \hat{A}_s)$ | local area of overlap with respect to equal number of points $i = 1, ..., M$ |
| $EP_s, EP'_s, \widehat{EP}_s$ | (partial) Explanatory Power of the respective overlap $O_s \equiv O(A'_s, \hat{A}_s)$ |
| $\overline{EP}, \overline{EP}', \overline{\widehat{EP}}$ | average of $EP_s$ values for or all overlaps, i.e. overall explanatory power of the modeled data series compared to observations |

## AI: Deduction of equations (9) – (13)

The expression (9c) for the INV slope immediately follows by putting $\lambda = 0$ in equation (8). Dividing (8) by $\lambda$ first and considering the limit $\lambda \to \infty$ similarly gives expression (9a) for the OLS slope. For the special RMA condition $\lambda = S^2_{y'}/S^2_{x'}$ equation (8) simplifies to $\hat{c}^2 = \lambda$ which gives (9b) as its positive root.

Equation (11) directly follows from (9a-c) using Pearson's correlation coefficient $R = S_{x'y'}/(S_{x'} \cdot S_{y'})$ as abbreviation. For $0 \leq R < 1$ the set of inequalities (10) is obtained directly from (11), while for $R = 1$ all three expressions in (10) are equal.

The general slope $\hat{c}_{\mathrm{EVM}}(\lambda) \equiv \hat{c}$ expressed in (7a) is the positive solution of equation (8); according to our assumption $S_{x'y'} > 0$ the other solution is always negative. Equation (8) can then be eliminated for $\lambda$ which gives (after canceling $\hat{c} \cdot S_{x'y'}$ in numerator and denominator)

$$\lambda = \frac{\hat{c} \cdot S^2_{y'} - \hat{c}^2 \cdot S_{x'y'}}{\hat{c} \cdot S^2_{x'} - S_{x'y'}} = \frac{\dfrac{S^2_{y'}}{S_{x'y'}} - \hat{c}}{\dfrac{S^2_{x'}}{S_{x'y'}} - \dfrac{1}{\hat{c}}} = \frac{\hat{c}_{\mathrm{INV}} - \hat{c}}{(\dfrac{1}{\hat{c}_{\mathrm{OLS}}} - \dfrac{1}{\hat{c}})} \qquad (A1)$$

as the first part of equation (12). The second and third part can be obtained by replacing $\hat{c}_{\mathrm{INV}}$ by $\hat{c}_{\mathrm{RMA}}$ or $\hat{c}_{\mathrm{OLS}}$ with the help of equation (11).

The variance of the modeled time series $\hat{y}'$ as the average sum of squares of deviation from the series mean is given as

$$S^2_{\hat{y}'} := \sum_{i=1}^{N}(\hat{y}'_i - \bar{y}')^2 = \sum_{i=1}^{N}(\hat{c} \cdot \hat{x}'_i + \hat{c}_0 - (\hat{c} \cdot \bar{x}' + \hat{c}_0))^2 = \hat{c}^2 \cdot \sum_{i=1}^{N}(x'_i - \bar{x})^2 = \hat{c}^2 \cdot S^2_{x'} \qquad (A2)$$

where we used the series properties from (4) as well as $\bar{\hat{y}}' = \bar{y}'$, hence for the overall ratio of modeled to observed variance we get

$$\frac{S_{\hat{y}'}^2}{S_{y'}^2} = \frac{\hat{c}^2 \cdot S_{x'}^2}{S_{y'}^2} = \frac{\hat{c}^2}{\hat{c}_{\text{RMA}}^2} \tag{A3}$$

as equation (13). One can immediately see that for fixed variances of $x'$ and $y'$ the variance ratio of $\hat{y}'$ and $y'$ is strictly increasing in $\hat{c}$ from minimal value $\hat{c}_{\text{OLS}}$ to maximal value $\hat{c}_{\text{INV}}$ and is smaller than one for noise ratios below RMA and larger than one above. It should be remarked that equation (A3) is also valid for any subsection of the time series, if means and variances are calculated accordingly.

**AII: Properties of the partial EP-ratio $Q_{EP}$ in dependence of elementary fluctuations**

Independent from the error noise conditions in the data it holds:

$$\hat{c}_{RMA} = \hat{c}_{OLS}/R = \hat{c}_{INV} \cdot R \text{ thus } \hat{c}_{OLS} \leq \hat{c}_{RMA} \leq \hat{c}_{RMA}$$

and: $R = S_{\hat{y}_{OLS}}/S_{y\prime}$; $1/R = S_{\hat{y}_{INV}}/S_{y\prime}$,

wherefore the following equation is valid:

$$\frac{R}{R} = \frac{S_{\hat{y}_{INV}} \cdot S_{\hat{y}_{OLS}}}{S_{y\prime}^2} = 1 \tag{A4}$$

leading to: $S_{y\prime}^2 \equiv S_{\hat{y}_{INV}} \cdot S_{\hat{y}_{OLS}}$

Under RMA conditions (i.e. $\lambda = S_{y\prime}^2/S_{x\prime}^2$) the following equation becomes valid:

$$1 = \frac{S_{\hat{y}_{INV}} \cdot S_{\hat{y}_{OLS}}}{S_{y\prime}^2} = \frac{\hat{c}_{INV} \cdot S_{x\prime} \cdot \hat{c}_{OLS} \cdot S_{x\prime}}{S_{y\prime}^2} := \frac{\hat{c}_{INV} \cdot S_\varepsilon \cdot \hat{c}_{OLS} \cdot S_{\varepsilon\prime}}{S_\delta^2} = \frac{\hat{c}_{RMA}^2 \cdot S_\varepsilon^2}{S_\delta^2} = \frac{S_{\hat{\delta}}^2}{S_\delta^2} = 1$$

The ratio

$$Q_{EP} := \frac{\widehat{EP}(\hat{c}_{OLS})}{EP'(\hat{c}_{INV})} \text{ with } \widehat{EP}(\hat{c}_{OLS}) = \frac{O_{OLS}}{2 \cdot \hat{c}_{OLS} \cdot S_{x\prime}} \quad , \quad EP'(\hat{c}_{INV}) = \frac{O_{INV}}{2 \cdot S_{y\prime}}$$

can theoretically be determined for any part of the data (independent from the local extreme values), where the overlaps $O_{INV}$ and $O_{OLS}$ can also be expressed using the factors $(f_{INV}, f_{OLS})$ as functions of $S_{x\prime}$ and $S_{y\prime}$ respectively with $1 \geq (f_{INV}, f_{OLS}) \geq 0$:

$$\widehat{EP}(\hat{c}_{OLS}) = \frac{2 \cdot f_{OLS} \cdot S_{y\prime}}{2 \cdot \hat{c}_{OLS} \cdot S_{x\prime}} \quad \text{(A5a)}, \quad EP'(\hat{c}_{INV}) = \frac{2 \cdot f_{INV} \cdot S_{x\prime} \cdot \hat{c}_{INV}}{2 \cdot S_{y\prime}} \tag{A5b}$$

It follows:

$$Q_{EP} = \frac{\widehat{EP}(\hat{c}_{OLS})}{EP'(\hat{c}_{INV})} = \frac{f_{OLS}}{f_{INV}} \cdot \frac{S_{y\prime} \cdot S_{y\prime}}{\hat{c}_{OLS} \cdot S_{x\prime} \cdot \hat{c}_{INV} \cdot S_{x\prime}} = \frac{f_{OLS}}{f_{INV}} \cdot \frac{S_{y\prime}^2}{S_{\hat{y}_{OLS}} \cdot S_{\hat{y}_{INV}}} \tag{A6}$$

Due to: $1 \geq R = S_{\hat{y}_{OLS}}/S_{y\prime}$, $\hat{c}_{OLS} \cdot S_{x\prime} \leq S_{y\prime}$, holds for all data conditions wherefore the bandwidths corresponding with $\hat{c}_{OLS} \cdot S_{x\prime}$ always will be encompassed by the respective bandwidths of $S_{y\prime}$. It follows that $\widehat{EP}(\hat{c}_{OLS})=1$ and therefore in (A5a) $f_{OLS} = R$. In analogy hereto it follows for (A5b) and $EP'(\hat{c}_{INV})$: $f_{INV} = R$. Thus Q$_{EP}$ (A6) always has to be 1 if $\hat{c}_{OLS}$ and $\hat{c}_{INV}$ of the respective dataset are used for the calculation of $EP'$ and $\widehat{EP}$, for instance over the complete length of the series and as long as EVM conditions (white noise) are fulfilled.

If now considering the elementary fluctuations under the assumption that the noiseless signal has different spectral characteristics than white noise, the arrangement of the sequences along the local extremes of the data will mainly occur due to the extremes of the noise (see also section 3.1). Then, for each sequence s=1,…,M the ratio

$$\left(\frac{\widehat{EP}(\hat{c}_{OLS})}{EP'(\hat{c}_{INV})}\right)_s = \left(\frac{f_{OLS}}{f_{INV}} \cdot \frac{S_{y\prime}^2}{\hat{c}_{OLS} \cdot \hat{c}_{INV} \cdot S_{x\prime}^2}\right)_s = \left(\frac{f_{OLS}}{f_{INV}} \cdot \frac{f_{y\prime} \cdot S_\delta^2}{\hat{c}_{RMA}^2 \cdot f_{x\prime} \cdot S_\varepsilon^2}\right)_s$$

will be an approximated local representation of $\lambda$ (see also A9). In contrast to a calculation of the ratio over the complete length of the compared series (A5), this ratio will only become 1 under RMA conditions because only under these conditions $\hat{c}_{RMA}^2 \cdot S_\varepsilon^2 = S_\delta^2$. This means, that any deviation from one will reflect the deviation from RMA error noise conditions. Due to the small sample size of each fluctuation which results in local violations of white noise conditions, the local ratios will have a certain error, but averaged over all fluctuations these errors should theoretically diminish wherefore Q$\overline{EP}_{max}$ should be a good estimate on the deviation from RMA conditions.

Even if the contributions of the local standard deviations from the noiseless data ($f_{x'}$ and $f_{y'}$) distort the amplitudes of the elementary fluctuations these distortions would diminish in direction of RMA conditions (with $f_{x'} = f_{y'}$ at the RMA point) if assuming a linear deterministic relationship between x and y and as long as the arrangement of the elementary fluctuations mainly occurs due to the structures of the error noise.

### AIII: Using Q$\overline{EP}_{max}$ to estimate c

The ratio Q$\overline{EP}_{max} = \frac{max(\overline{EP})}{max(\overline{EP'})}$ from equation (19) can be used to estimate the ratio $\lambda$ of noise variances and therefore to provide an estimate $\tilde{\hat{c}}_Q$ of the slope. According to definition (15a) in each (short) fluctuation it holds that

$$A'_s := 2 \cdot S_{y'_s} \approx 2 \cdot S_\delta \quad , \quad \hat{A}_s := 2 \cdot S_{\hat{y}'_s} \approx 2 \cdot \hat{c} \cdot S_\varepsilon \quad , \quad s = 1, ..., M \quad . \tag{A6}$$

Putting this into the definitions (16) of the partial explanatory powers, we get

$$\frac{\widehat{EP}_s(\hat{c})}{EP'_s(\hat{c})} \cdot \hat{c} = \frac{A'_s}{\hat{A}_s} \cdot \hat{c} = \frac{2 \cdot S_\delta}{2 \cdot \hat{c} \cdot S_\varepsilon} \cdot \hat{c} = \sqrt{\lambda} = \text{const} \tag{A7}$$

for any $\hat{c} = \hat{c}_{it}$ applied in the iteration. It means, that the ratio (A7) stays constant with $\hat{c}$ within the formal variation of $\hat{c}$ from $\hat{c}_{OLS}$ to $\hat{c}_{INV}$ under idealized EVM conditions, because the error noises in the data do not vary. This holds for each elementary fluctuation, as long as the noiseless data are not completely random values. Close to RMA conditions both numerator and denominator in the quotient of the left term in (A7) can be replaced by the maximum values taken for the extreme slopes $\hat{c}_{OLS}$ and $\hat{c}_{INV}$, respectively, while $\hat{c}$ is being replaced by $\hat{c}_{RMA}$. If averaged over all sequences s, this gives the approximation

$$\lambda \approx \hat{c}^2_{RMA} \cdot \left(\frac{\widehat{EP}(\hat{c}_{OLS})}{EP'(\hat{c}_{INV})}\right)^2 \approx \hat{c}^2_{RMA} \cdot Q\overline{EP}^2_{max} \tag{A8}$$

which becomes unbiased for the RMA case.

Hence Q$\overline{EP}_{max} \to 1$ can also be used as convergence criterion to RMA conditions. If we put the abbreviations (9a-c) into equation (7a) we directly get

$$\tilde{\hat{c}}_Q = \hat{c}_{INV} \cdot \frac{1 - Q\overline{EP}^2_{max}}{2} + \sqrt{\hat{c}_{RMA} \cdot Q\overline{EP}^2_{max} + \hat{c}^2_{INV} \cdot \frac{(1 - Q\overline{EP}^2_{max})^2}{4}} \tag{A9}$$

as an estimate (20) of the slope via $Q\overline{EP}_{max}$. Of course, putting $Q\overline{EP}_{max} = 1$ in equation (20) gives $\tilde{\hat{c}}_Q = \hat{c}_{RMA}$, while $Q\overline{EP}_{max} = 0$ gives $\tilde{\hat{c}}_Q = \hat{c}_{INV}$ and $Q\overline{EP}_{max} \to \infty$ gives $\tilde{\hat{c}}_Q = \hat{c}_{OLS}$.